\documentclass{article}
\usepackage{ijcai23}
\usepackage{times}
\usepackage{soul}
\usepackage{url}
\usepackage[hidelinks]{hyperref}
\usepackage[utf8]{inputenc}
\usepackage[small]{caption}
\usepackage{graphicx}
\usepackage{amsmath}
\usepackage{amsthm}
\usepackage{booktabs}
\usepackage{algorithm}
\usepackage{algorithmic}
\usepackage[switch]{lineno}
\title{Continuous-Time Graph Learning for Cascade Popularity Prediction}

\author{
Xiaodong Lu
\and
Shuo Ji\and
Le Yu \and
Leilei Sun \and
Bowen Du \and
Tongyu Zhu\thanks{Corresponding Author}
\affiliations
SKLSDE Lab, Beihang University, Beijing 100191, China
\emails
\{xiaodonglu, jishuo, yule, leileisun, dubowen, zhutongyu\}@buaa.edu.cn
}

\newcommand{\figref}[1]{Figure \ref{#1}}
\newcommand{\tabref}[1]{Table \ref{#1}}
\newcommand{\secref}[1]{Section \ref{#1}}

\newcommand{\equref}[1]{Equation (\ref{#1})}
\newcommand{\citet}[1]{\citeauthor{#1}\shortcite{#1}}
\usepackage{multirow}
\usepackage{multicol}
\usepackage{bm}
\usepackage{amsmath}
\usepackage{xcolor}
\usepackage{subcaption}
\begin{document}
\maketitle

\begin{abstract}
Information propagation on social networks could be modeled as cascades, and many efforts have been made to predict the future popularity of cascades. However, most of the existing research treats a cascade as an individual sequence. Actually, the cascades might be correlated with each other due to the shared users or similar topics. Moreover, the preferences of users and semantics of a cascade are usually continuously evolving over time. In this paper, we propose a continuous-time graph learning method for cascade popularity prediction, which first connects different cascades via a universal sequence of user-cascade and user-user interactions and then chronologically learns on the sequence by maintaining the dynamic states of users and cascades. Specifically, for each interaction, we present an evolution learning module to continuously update the dynamic states of the related users and cascade based on their currently encoded messages and previous dynamic states. We also devise a cascade representation learning component to embed the temporal information and structural information carried by the cascade. Experiments on real-world datasets demonstrate the superiority and rationality of our approach.
\end{abstract}

\section{Introduction}
% cascade definition & example -> prediction importance 
% In social network, the process of the information dissemination 

The information propagation, aka the information cascade, is ubiquitous on online social networks, which records human behaviors in posting and accessing information. For example, on Twitter, a tweet posted by a user may disseminate to other users, and such retweeting behaviors between users can be denoted as an information cascade. Predicting the popularity of such information cascades could help people understand the information propagation better and is crucial for numerous applications such as viral marketing \cite{DBLP:journals/tweb/LeskovecAH07}, scientific impact qualification \cite{DBLP:journals/corr/GuoS14b} and item recommendation \cite{DBLP:conf/kdd/WuGGWC19}. %rumor control?

\begin{figure}[!h]
    \centering
    \includegraphics[width=\columnwidth]{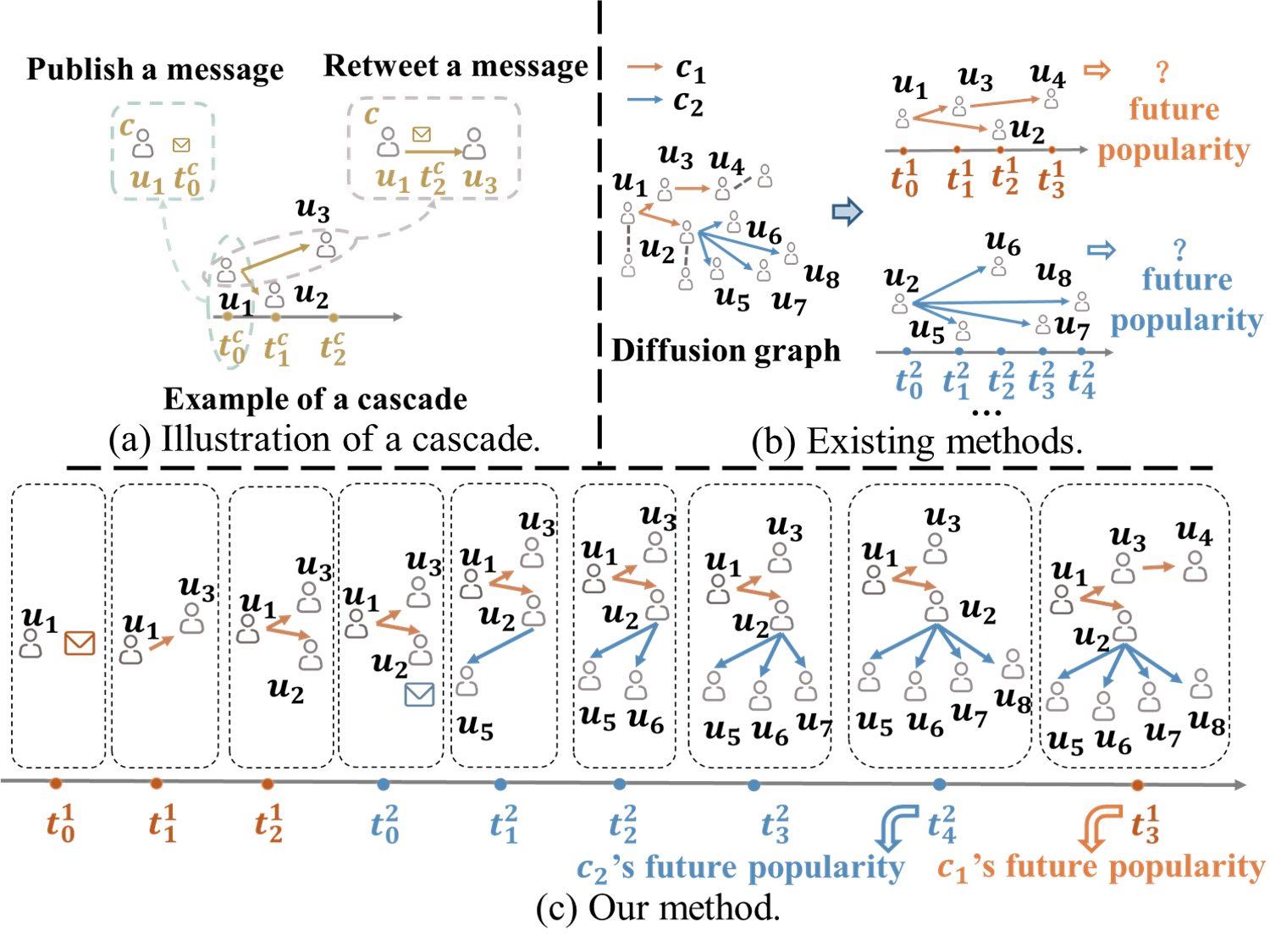}
    % \caption{Different from existing methods which mainly learn on each cascade's own sequence, our method learns continuously evolving representations of cascades and users collaboratively, which can exploit the correlations of different cascades and capture the continuous dynamic preference of users.}
    \caption{Different from existing methods which mainly learn on each cascade's own sequence, our method learns continuously evolving representations of cascades and users collaboratively, which can model the correlation between cascades and continuously dynamic user preferences. 
    }
    \label{fig:cascade}
\end{figure}

% methods list
Up to now, lots of attempts have been made on this problem. In the early stage, researchers extracted manual features to represent a cascade \cite{DBLP:conf/www/ChengADKL14,DBLP:journals/cacm/SzaboH10}. However, these methods are based on hand-designed features with a relatively large number of human efforts and may lose useful information during the diffusion behavior of a cascade. Different from feature-based methods, some researchers considered the cascade as a diffusion sequence and employed sequential models like recurrent neural networks to capture the evolution pattern of cascades \cite{DBLP:conf/aaai/LiaoXLHLL19,DBLP:conf/cikm/CaoSCOC17,DBLP:conf/ijcai/YangTSC019}, while the structural information within the cascade has not been fully exploited yet. Recently, graph representation learning methods were introduced to further improve the prediction performance \cite{DBLP:conf/www/LiMGM17,DBLP:conf/icde/Chen0ZTZZ19,DBLP:conf/sigir/ChenZ0TZZ19,casflow,tempcas}. These methods utilized the social network and cascade graph to learn the structural and temporal information within each cascade. 

Although insightful, most of the existing methods solely predict the popularity of each cascade within its own sequence, see \figref{fig:cascade} (b). We argue that there are two essential factors that are not well considered by previous methods. Firstly, different cascades can be correlated with each other because of the shared users or similar semantics. For example, in \figref{fig:cascade} (b), we aim to predict the popularity of cascade $c_1$ and existing methods can only learn from the propagation sequence of $c_1$ itself. However, if we additionally consider cascade $c_2$, we can find that both $c_1$ and $c_2$ contain user $u_2$ which seems to be a popular user with many retweets, and this is helpful for predicting the popularity of $c_1$.
% For example, a user's next forward message will relate to message he forward this time \cite{DBLP:conf/ijcai/WangZWYZ18}. 
Secondly, the states of users are often evolving in a continuous manner (e.g., a user is likely to gradually change his interests according to the information he/she received from the social network at different times), which cannot be captured by existing methods either.
% However, previous methods that learn on the network snapshots fail to capture these two properties.
% . Though some existing methods model preference of users, which are considered statically. This is inconsistent with the realistic scenarios where users' preferences may evolve. For example, a scholar may change his research interests and thus make the cascades that he participates in be different from before.

To tackle the above issues, we propose a \textbf{C}ontinuous-\textbf{T}ime graph learning method for \textbf{C}ascade \textbf{P}opularity prediction, namely CTCP. To model the correlation between cascades, we first combine all cascades into a dynamic diffusion graph as shown in \figref{fig:cascade} (c), which can be considered as a universal sequence of diffusion behaviors (i.e, user-cascade and user-user interactions). Then, we propose an evolution learning module to chronologically learn on each diffusion behavior by maintaining a dynamic representation for each user and cascade that evolves continuously as the diffusion behavior happens. When a diffusion behavior happens, this module first encodes the information of a diffusion behavior into a message and then fuses the dynamic representations of related users and cascades with the generated message. Next, a cascade representation learning model is proposed to generate the static and dynamic cascade embeddings by aggregating user representations from both temporal and structural perspectives. Based on the generated embeddings, the prediction module finally computes the cascade popularity. The main contributions of the paper are summarized as follows.
\begin{itemize}
    \item Different from previous methods that only learn from the own sequence of each cascade, we propose a continuous-time graph learning method to explore the correlations of different cascades by a dynamic diffusion graph and explicitly learn the dynamic preferences of users in the network.
    \item We maintain dynamic representations for users and cascades and design an evolution learning module to encode the information of a diffusion behavior into a message and continuously update the dynamic representations by fusing the previous dynamic representations with the message in a recurrent manner.
    \item A cascade representation learning module is proposed to capture the temporal and structural information of a cascade, which leverages the sequence and graph structure to aggregate representations of users.
\end{itemize}

% The remaining of this paper is organized as follows: we first present the formalization of the studied problem, then provide details of the proposed method. Experiments have been conducted on three real-world data sets. Finally, we introduce the related literature and conclude this research.

\section{Related Work}
\subsection{Cascade Popularity Prediction} 
The cascade popularity prediction problem aims at predicting the future size of an information cascade. Many efforts have been paid to this problem. In the early stage, researchers represented a cascade as some handcrafted features such as content features and user  attributes \cite{DBLP:conf/www/ChengADKL14,DBLP:journals/cacm/SzaboH10}.  
% For example, \citet{DBLP:conf/www/ChengADKL14} used the average out degrees of cascade graph and the average time interval between retweets as parts of the cascade features. 
However, these methods need a relatively large number of human labor to design or select and  have limited generalization ability. 

Different from the feature-based methods, some researchers considered the cascade as a diffusion sequence of users and employed the sequence-based model to learn the evolution pattern of a cascade \cite{DBLP:conf/aaai/LiaoXLHLL19,DBLP:conf/cikm/CaoSCOC17,DBLP:conf/ijcai/YangTSC019}. For example, \citet{DBLP:conf/cikm/CaoSCOC17} utilized the Gated Recurrent Unit (GRU) to learn a path-level representation and aggregate path representation into cascade representation by different learnable weights. Though the sequence-based methods achieve considerable performance, the structural information of cascades is not well explored.

To fully utilize the temporal and structural information within cascades, some graph-based methods have been proposed \cite{DBLP:conf/www/LiMGM17,DBLP:conf/icde/Chen0ZTZZ19,DBLP:conf/sigir/ChenZ0TZZ19,casflow,tempcas}, which modeled a single cascade as a graph evolving with time and leveraged the graph representation learning method to learn cascade representations from cascade graphs. However, these methods predict the popularity of each cascade separately and thus neglect the correlation between cascades. Some recent methods model the evolution of multiple cascades by considering it as a sequence of graph snapshots sampled at regularly-spaced times \cite{DBLP:conf/sigir/00040LSL0WSYA21,DBLP:conf/aaai/0004RZLY22}, but these methods discrete the continuous timestamps into several regularly-spaced time steps and thus can not model the continuous evolution of user preferences. Moreover, these methods may have high memory overhead, because it needs to load a whole graph snapshot at one time. 

In summary, although insightful, existing methods have not well addressed the issues of correlation between cascades and the dynamic evolution of user preferences.

\subsection{Graph Representation Learning}
In recent years, the Graph Neural Network (GNN) has achieved superior performance on graph representation learning. To be better in accordance with real-world scenarios, some researchers have further designed heterogeneous GNNs and dynamic GNNs \cite{DBLP:journals/jmlr/KazemiGJKSFP20,DBLP:journals/corr/abs-2011-14867,DBLP:conf/kdd/HuangSDLL021}. For example, \citet{DBLP:conf/esws/SchlichtkrullKB18} focused on the relation learning in knowledge graphs.
\citet{DBLP:conf/www/WangJSWYCY19} and \citet{DBLP:conf/www/HuDWS20} studied heterogeneous graphs based on meta-paths and the attention mechanism. Some researchers \cite{DBLP:conf/aaai/ParejaDCMSKKSL20,DBLP:conf/wsdm/SankarWGZY20} treated a dynamic graph as a sequence of snapshots, while others \cite{DBLP:conf/iclr/XuRKKA20,DBLP:conf/cikm/ChangLW0FS020,DBLP:journals/corr/abs-2006-10637} modeled each dynamic graph as a temporal graph or a sequence of events.

In this paper, we investigate the correlation between cascades and the dynamic user preferences by considering the evolution of cascades as a continuous-time graph.

\section{Problem Formulation}
\textbf{Cascade.} Given a set of users $\mathcal U$, a cascade $c$ records the diffusion process of a message $m$ among the users $\mathcal U$. Specifically, we use a chronological sequence  $g^c(t)=\{(u_i^c,v_i^c,t_i^c)\}_{i=1,...,|g^c(t)|}$ to represent the growth process of cascade $c$ until time $t$, where $(u_i^c,v_i^c,t_i^c)$ indicates that $v_i^c$ forwards the message $m$ from $u_i^c$ (or we can say that $v_i$ participates in cascade $c$ through $u_i$). In addition, we use $(u_0^c,t_0^c)$ to denote that $u_0^c$ publishes the message $m$ at $t_0^c$ (or we can say that $u_0^c$ begins cascade $c$ at $t_0^c$).

\textbf{Diffusion Graph.} Based on the above definitions, we use  the diffusion graph $\mathcal G_d^t=\{(u_i,v_i,c_i,t_i)|t_i<t\}$  to denote the diffusion process of all cascades until $t$. Here $(u_i,v_i,c_i,t_i)$ is a diffusion behavior representing that $v_i$ participates in cascade $c_i$ through $u_i$ at $t_i$. The diffusion graph $\mathcal G_d^t$ can be considered as a chronological sequence of diffusion behaviors as shown in \figref{fig:cascade} (c).

\textbf{Cascade Prediction.} Given a cascade $c$ begins at $t^c_0$, after observing it for time $t_o$, we want to predict its incremental popularity $\Delta P_c = |g^c(t_0^c + t_p)| - |g^c(t_0^c+t_o)|$ from $t_0^c + t_o$ to $t_0^c+t_p$, where $t_p >> t_o$ is the prediction time.

Most of the previous methods consider the task as a single cascade prediction problem, that is, learning a function $f:g^c(t_0^c+t_o) \rightarrow \Delta P_c$ that predicts the incremental popularity of  a cascade only based on its own historical observation. However, the collaborative signals between the cascades are ignored, which motivates us to design our new method to consider other cascades when predicting the incremental popularity of a cascade. Specifically, we learn a function $f:g^c(t_0^c+t_o) \times {G}_d^{t_0^c+t_o}  \rightarrow \Delta P_c$ which not only considers the information of a single cascade but also takes the historical diffusion on the social network into account.

\section{Methodology}
\begin{figure*}[htbp]
    \centering
    \includegraphics[width=0.91\linewidth]{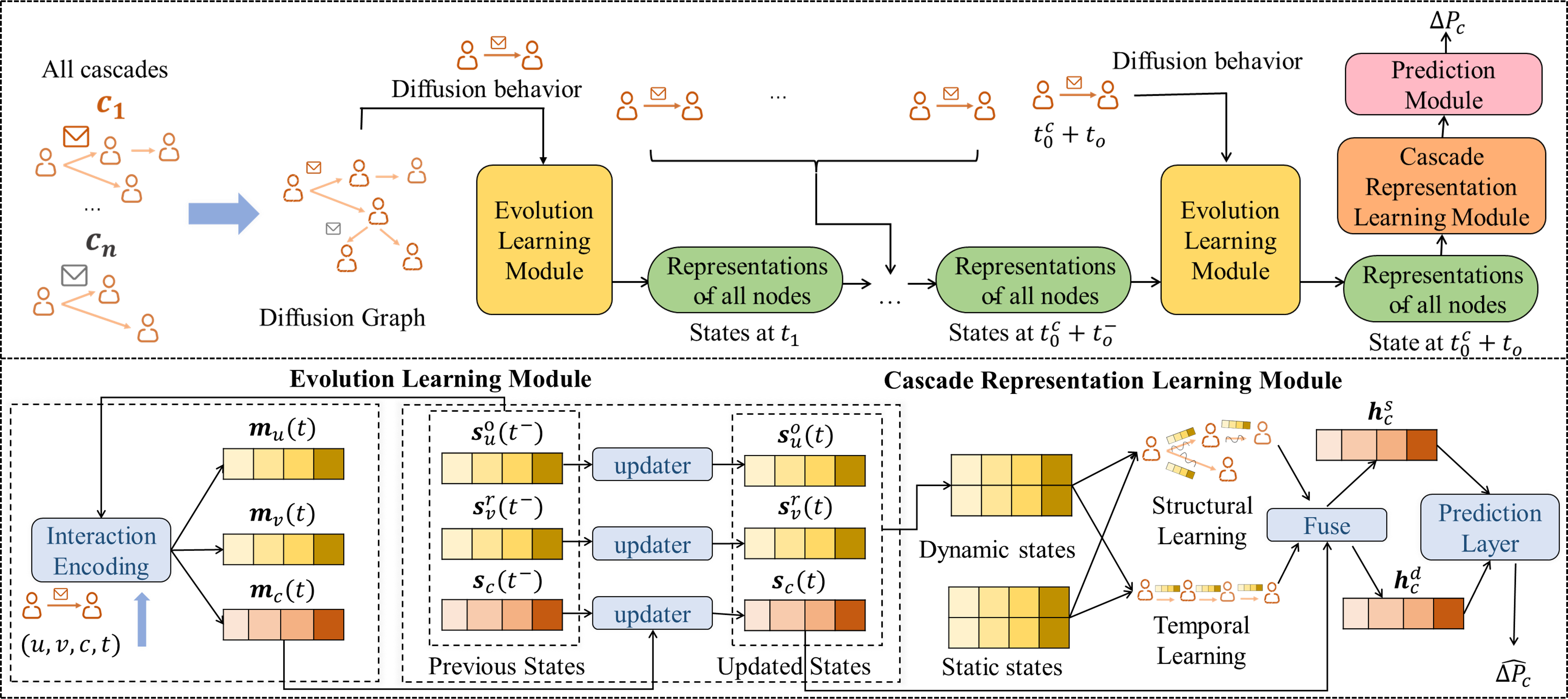}
    \caption{Framework of the proposed method. It consists of an evolution learning module to model the dynamics of user preferences and the correlation between cascades, a cascade representation learning module to capture the temporal and structural information within a cascade, and a prediction module to give future popularity. The popularity of cascade is predicted according to the most recent representations. }
    \label{fig:framework}
\end{figure*}

As shown in \figref{fig:framework}, we first consider all cascades into a chronological sequence of diffusion behaviors (i.e., the diffusion graph). Then we learn on each diffusion behavior sequentially, where we maintain continuously evolving representations for cascades and users to explore the dynamic preference of users and the correlation between cascades. During the sequential learning process, whenever the observation time $t_o+t^c_0$ of a cascade $c$ is reached, we predict its incremental popularity $\Delta P_c$. 

Specifically, our method consists of three components: 1) Evolution learning module maintains dynamic states (i.e., the dynamic representation) for users and cascades, which models cascades in diffusion behavior level (micro).  2) Cascade representation learning module generates the embeddings of cascades by aggregating user representations from different perspectives, which models cascades in a diffusion structure level (macro). 3) Prediction module gives the prediction of the incremental popularity of cascades.

\subsection{Evolution Learning Module}
\textbf{Dynamic States.}
 We first introduce the dynamic states for users and cascades. From the perspective of information diffusion, there are two roles of a user: originator and receiver. For example, in a diffusion behavior $(u,v,c,t)$, user $u$ acts as the originator of the message, and user $v$ acts as the receiver of the message. Thus, we maintain two types of dynamic states $\bm{s}^o_u(t)$ and $\bm{s}^r_u(t)$ for a user to describe his originator role and receiver role respectively. Besides, we maintain a dynamic state $\bm{s}_c(t)$ for every cascade $c$ to memorize its diffusion history, which can help users get information from previous participating users in the cascade. The above dynamic states are initialized to zero vectors and learned from the global diffusion behavior sequence.

\textbf{Dynamic State Learning.}
When a diffusion behavior $(u,v,c,t)$ happens, the dynamic states of the corresponding users and cascade should be updated. Naturally, the behaviors of a user (cascade) can be considered as a sequence and sequential models like the recurrent neural network can be employed to learn dynamic states from the sequence of a user (cascade). In addition to the own behaviors of a user (cascade), there are also global dependencies needed to be considered. For example, when a user $u$ participates in a diffusion behavior $(u,v,c,t)$, he may also be influenced by the users who previously participated in the cascade $c$. To this end, we employ a recurrent neural network $f_r(\cdot)$ to update the dynamic states of users and cascades globally. Specifically, when a diffusion behavior $(u,v,c,t)$ happens, we update the states of $u,v,c$ by $f_r(\cdot)$.  The updating process consists of two steps: interaction encoding and state updating. In the interaction encoding, we encode the information of diffusion behavior $(u,v,c,t)$ and generate message $\bm{m}_u(t)$, $\bm{m}_v(t)$, $\bm{m}_c(t)$ for $u$, $v$ and $c$ to guide the subsequent state updating process. Assuming the state of $u$ before $t$ is $\bm{s}^{o}_u(t^-)$, we generate message representation for user $u$ by the following mechanism, 
\begin{align}
    \bm{f}^t_u &= [\cos{w_1^r\Delta t_u} ,\cos{w_2^r\Delta t_u},...,\cos{w_n^r\Delta t_u}], \\
    \bm{m}_u(t) &= \sigma(\bm{W}^{r}[\bm{s}^{o}_u(t^-)||\bm{s}^{r}_v(t^-)||\bm{s}_c(t^-)||\bm{f}^t_u] + \bm{b}^r), 
\label{eq:message}
\end{align}
where $||$ is the concatenation operation, $\Delta t_u$ is the time interval since the last updating of users $u$ (i.e., $\Delta t_u = t - t_u^-$ and $t_u^-$ is the last time where $u$ was updated), and $\bm{f}^t_u$ is the temporal feature learned from a series of cosine basis functions. 

After generating the message representation, we fuse the old dynamic state $\bm{s}_u^o(t^-)$ with the message representation $\bm{m}_u(t)$ to get the updated states $\bm{s}_u^o(t)$ by GRU \cite{DBLP:conf/emnlp/ChoMGBBSB14}, 
\begin{equation}
\resizebox{.9\linewidth}{!}{
    $
    \displaystyle
    \begin{split}
        \bm{g}_i &= \sigma(\bm{W}_{i,s} \bm{s}_u^o(t^-)+ \bm{W}_{i,m}\bm{m}_u(t)  + \bm{b}_i), \\
        \bm{g}_f &= \sigma(\bm{W}_{f,s} \bm{s}_u^o(t^-)+ \bm{W}_{f,m}\bm{m}_u(t)  + \bm{b}_f), \\
        \bm{\hat{s}}_u^o(t) &= \tanh(\bm{W}_{m} \bm{m}_u(t) + \bm{g}_i \odot(\bm{W}_{s} \bm{s}_u^o(t^-)+\bm{b}_s) + \bm{b}), \\
        \bm{s}_u^o(t) &= \bm{g}_f \odot \bm{s}_u^o(t) + (1-\bm{g}_f)\odot \bm{\hat{s}}_u^o(t^-),
    \end{split}$}
    \label{eq:update}
\end{equation}
The updating process of user $v$ and cascade $c$ is the same as user $u$ in addition to different learnable parameters. 

\subsection{Cascade Representation Learning Module}
In this module, we generate embeddings for cascades by aggregating representations of participating users. Specifically, we learn the temporal and structural characteristics of a cascade by leveraging the diffusion sequence and cascade graph to aggregate representations of users respectively. Besides the dynamic states $\bm{s}_u^o(t)$ and $\bm{s}_u^r(t)$ of users, we also introduce the static state $\bm{s}_u$ to represent the static preference of a user $u$. The static state is initialized randomly and learnable during the training process. 

\textbf{Temporal Learning.} Given a cascade $c$, we organize it as a diffusion sequence of users $U_c = {(u_1,t_1),(u_2,t_2),...,(u_n,t_n)}$ where $(u_i,t_i)$ indicates that user $u_i$ participate in the cascade $c$ at $t_i$ after the publication of the cascade. The target of this module is to learn the temporal pattern from the diffusion sequence such as the short-term outbreak of user participation. The direct way to learn the temporal pattern is feeding participating users' representations sequentially into a recurrent neural network, however, it may neglect the time information in the diffusion sequence since it can not distinguish users participating at different times. Inspired by the position embedding technics \cite{DBLP:conf/nips/VaswaniSPUJGKP17}, we divide the observation time $t_o$ into $n_t$ slots $[0,\frac{t_0}{n_t}), [\frac{t_0}{n_t},2\frac{t_0}{n_t}) ..., [(n_t-1)\frac{t_0}{n_t},t_o)$ and preserve a learnable embedding $\bm{e}^t_i$ for every time interval $[i\frac{t_0}{n_t},(i+1)\frac{t_0}{n_t})$ to distinguish users that participate in the cascade at different time. Besides, we also introduce another learnable parameter $\bm{e}^p$ to strengthen the path information, where $\bm{e}^p_i$ is a position embedding for the ith participating users. We get the user embedding $\bm{z}^s_{u_i}$ by first adding these two embeddings to the states of users and then feeding the sequence of user embeddings to the Long Short-Term Memory (LSTM) \cite{lstm} to get the cascade temporal embedding $\bm{h}_{c,t}^s$, that is, 
\begin{align}
        \bm{h}_{c,t}^s &= LSTM^s([\bm{z}^s_{u_1},\bm{z}^s_{u_2},...,\bm{z}^s_{u_n}]), \\
        \bm{z}^s_{u_i} &=  \bm{s}_{u_i} + \bm{e}^t_{[t_i]} + \bm{e}^p_i,
        \label{eq:temporal}
\end{align}
where $[t_i]$ is the time slot that $t_i$ belongs to, i.e, $[t_i]*\frac{t_0}{n_t} \leq t_i < ([t_i]+1)*\frac{t_0}{n_t}$. Here the superscript of $\bm{h}_{c,t}^s$ means it is the static temporal representation and we also generate a dynamic temporal representation $\bm{h}_{c,t}^d$ by the above equation except using different LSTM parameters and user representations (i.e., dynamic states).

\textbf{Structural Learning.} Besides the order and time interval of participating users, the cascade graph also plays an important role in popularity prediction. For example, a deeper cascade may get more popularity since it influences the users who are far away from the original user \cite{DBLP:conf/www/ChengADKL14}. The cascade graph can be considered as a directed acyclic graph (DAG) with a root node (the user who posts the message), where a path from the root node to other nodes represents a diffusion process of a message in the social network. Though graph neural networks like GCN can learn graph structure, it may be difficult to model deep cascade paths \cite{tempcas}. Inspired by \cite{DBLP:conf/acl/TaiSM15,DBLP:conf/kdd/DucciKF20}, we employ a modified LSTM and aggregate representations of users on the cascade graph along the direction of information flow. Formally, let $S(u)$ and $T(u)$ be the users that $u$ receives messages from and sends messages to, i.e., there are edges pointing from $S(u)$ to $u$ and $u$ to $T(u)$. Then we employ the following mechanism to propagate the information from root nodes to leaf nodes. 
\begin{equation}
\begin{split}
    \centering
    \tilde{\bm{h}}^s_{u,\uparrow} &= \sum_{v \in S(u)} \bm{h}^s_{v,\uparrow}, \\
    \bm{i}^s_{u,\uparrow} &= \sigma(\bm{W}^s_{i,\uparrow}[\bm{s}_u||\tilde{\bm{h}}^s_{u,\uparrow}]+\bm{b}^s_{i,\uparrow}), \\
    \bm{f}^s_{uv,\uparrow} &= \sigma(\bm{W}^s_{f,\uparrow}[\bm{s}_u||\bm{h}^s_{v,\uparrow}]+\bm{b}^s_{f,\uparrow}), \\
    \bm{o}^s_{u,\uparrow} &= \sigma(\bm{W}^s_o[\bm{s}_u||\tilde{\bm{h}}^s_{u,\uparrow}]+\bm{b}^s_{o,\uparrow}), \\
    \bm{g}^s_{u,\uparrow} &= \tanh(\bm{W}^s_{g,\uparrow}[\bm{s}_u||\tilde{\bm{h}}^s_{u,\uparrow}]+\bm{b}^s_{g,\uparrow}), \\
    \bm{c}^s_{u,\uparrow} &= \bm{i}^s_{u,\uparrow} \odot \bm{g}^s_{u,\uparrow} + \sum_{v \in S(u)}\bm{f}^s_{uv,\uparrow}\odot \bm{c}^s_{v,\uparrow} \\
    \bm{h}^s_{u,\uparrow} &= \bm{o}^s_{u,\uparrow} \odot \tanh(\bm{c}^s_{u,\uparrow}), 
\end{split}
\label{eq:structural}
\end{equation}
After propagating the information in the graph, we sum the leaf nodes' representations to get the cascade embedding $\bm{h}^s_{c,\uparrow} = \sum \bm{h}^s_{leaf,\uparrow}$. Besides, we reverse the edge direction of the cascade graph and generate another cascade representation $\bm{h}^s_{c\downarrow}$ from leaf to root. Finally, we concatenate the $\bm{h}^s_{c,\uparrow}$ and $\bm{h}^s_{c,\downarrow}$ and feed it to an MLP to get the final structural representation $\bm{h}^s_{c,s}$. Here the superscript of $\bm{h}^s_{c,s}$ represents the static structural embedding of the cascade as in the temporal learning module. We also generate the dynamic representation $\bm{h}^d_{c,s}$ by the same mechanism in \equref{eq:structural} except using different parameters and user representations.

\textbf{Embedding Fusion.} In this module, we fuse the temporal embedding and structural embedding into a cascade embedding. For static embedding $\bm{h}^s_{c,t}$ and $\bm{h}^s_{c,s}$, we get the merged embedding $\bm{h}^s_c$ by concatenating them and then feed it into an MLP. The merge process of the dynamic embedding is slightly different from that of the static, where we split the participating users into two parts: the users $u$ and $v$ participating in the last diffusion $(u,v,c,t)$ of a cascade $c$ and others. The last two users $u,v$'s dynamic states are used to merge with the dynamic cascade state $\bm{s}_c(t)$ and the others are used to generate the temporal and structural embedding $\bm{h}^d_{c,t}$ and $\bm{h}^d_{c,s}$. The reason for this is that the last two users' dynamic states are updated from $\bm{s}^o_u(t^-),\bm{s}^r_v(t^-)$ to $\bm{s}^o_u(t),\bm{s}^r_v(t)$ by the updater in \eqref{eq:update} and this make the gradients can be propagated back to the updater through them, which makes them different from the dynamic states of other users. Formally, the merge process of the dynamic representation can be represented as
\begin{align}
    \bm{h}_c^d &= \sigma(\bm{W}_a[\bm{h}^d_{c,t}||\bm{h}^d_{c,s}||\tilde{\bm{h}}_c^d]), \\
    \tilde{\bm{h}}^d &= \sigma(\bm{W}_b[\tilde{\bm{s}_c}(t)||\bm{s}_u^o(t)||\bm{s}_v^r(t)]), \\
    \tilde{\bm{s}_c}(t) &= \bm{s}_c(t) + \bm{e}^g_{[t_0^c]},
\end{align}
where $t_o^c$ is the publication time of $c$ and $\bm{e}^g$ is another position embedding for publication time like \equref{eq:temporal}. 
\subsection{Prediction Module}
In this module, we give the prediction of incremental popularity by merging the prediction result from static embedding $\bm{h}_c^s$ and dynamic embedding $\bm{h}_c^d$.
\begin{align}
    \widehat{\Delta P}_c = \lambda f_{\textrm{static}}(\bm{h}_c^s) + (1-\lambda) f_{\textrm{dynamic}}(\bm{h}_c^d),
\end{align}
where the $f_{\textrm{static}}(\cdot)$ and $f_{\textrm{dynamic}}(\cdot)$ are two MLP functions and $\lambda$ is a hyperparameter to control the weight of static result and dynamic result. 

We use the Mean Squared Logarithmic Error (MSLE) as the loss function, which can be formulated as follows,
\begin{align}
    \mathcal{J}(\theta) = \frac{1}{n}\sum_{c} (log(\Delta P_c)-log(\widehat{\Delta P_c}))^2,
\end{align}
where $n$ is the number of training cascades.
\section{Experiments}
In this section, we conduct experiments on three datasets to evaluate the effectiveness of our approach.

\subsection{Descriptions of Datasets} 
We use three real-world datasets in the experiments, including the cascades in social platforms (Twitter and Weibo) and academic networks (APS).
\begin{itemize}
    \item \textbf{Twitter} \cite{weng2013virality} contains the tweets published between Mar 24 and Apr 25, 2012 on Twitter and their retweets during this period. Every cascade in this dataset represents the diffusion process of a hashtag. 
    \item \textbf{Weibo} \cite{DBLP:conf/cikm/CaoSCOC17} was collected on Sina Weibo which is one of the most popular Chinese microblog platform. It contains posts published on July 1st,2016 and their retweets during this period. Every cascade in this dataset represents the diffusion process of a post.
    \item \textbf{APS} \footnote{\url{https://journals.aps.org/datasets}} contains papers published on American Physical Society (APS) journals and their citation relationships before 2017. Every cascade in this dataset represents the process of obtaining citations for a paper. Following previous works \cite{DBLP:conf/cikm/CaoSCOC17}, transformation and preprocessing are taken to make paper citation prediction analogy to the retweet prediction.
\end{itemize}
% For the above datasets, we use the version provided in \cite{casflow}. 
Following \citet{casflow}, we randomly select $70\%$, $15\%$ and $15\%$ of the cascades for training,  validating and testing. 
% Some dataset settings and preprocessings follow previous works \cite{casflow,DBLP:conf/cikm/CaoSCOC17}: 
For data preprocessing, we set the observation window of a cascade to 2 days, 1 hour and 5 years on Twitter, Weibo and APS. For Weibo and Twitter, we predict cascades' popularity at the end of the dataset, while we predict cacades' popularity 20 years after its publication for APS. The cascades whose observed popularity $|c(t_0^c+t_o)|$ is less than 10 are discarded and for cascades whose $|c(t_0^c+t_o)|$ is more than 100, we only select the first 100 participants. Moreover, to ensure that there are adequate times for cascades to accumulate popularity and to avoid the effect of diurnal rhythm \cite{DBLP:conf/cikm/CaoSCOC17}, we select the cascades published before April 4th, published between 8:00 and 18:00, and published before 1997 on Twitter, Weibo and APS, respectively. The above preprocessing process also follows previous methods \cite{casflow,DBLP:conf/cikm/CaoSCOC17}.
% only the cascades published before April 4th on Twitter are selected and only the cascades published between 8:00 and 18:00 on Weibo are selected and only the cascades published before 1997 in APS are selected. 
\tabref{tab:dataset} shows the statistics of the datasets.

\begin{table}[!htbp]
    \small
    \centering
    \begin{tabular}{cccc}
    \toprule
    Datasets & \#Users & \#Cascades & \#Retweets \\ 
    \midrule
    Twitter & 199,005  & 19,718      & 602,253   \\ 
    \midrule
    Weibo   & 918,852  & 39,076      & 1,572,287   \\ 
    \midrule
    APS     & 218,323  & 48,575      & 939,686 \\  
    \bottomrule
    \end{tabular}
    \caption{Statistics of datasets.}
    \label{tab:dataset}
\end{table}

\subsection{Baselines} 
We compare our method with the following baselines, where the first two methods (i.e., XGBoost and MLP) additionally need hand-designed features (see details in \secref{sec:experimental_settings}):

\begin{itemize}
\item \textbf{XGBoost} belongs to the gradient boosting algorithm, which is a widely used machine learning method \cite{DBLP:conf/kdd/ChenG16}. 

\item \textbf{MLP} uses the multilayer perceptron to compute on the features of each cascade.

\item \textbf{DeepHawkes} \cite{DBLP:conf/cikm/CaoSCOC17} treats each cascade as multiple diffusion paths of users and learns sequential information of cascades through the GRU. 

\item \textbf{DFTC} \cite{DBLP:conf/aaai/LiaoXLHLL19} considers each cascade as a popularity count sequence and uses the Convolutional Neural Network (CNN), LSTM and attention mechanism to learn the cascade representation.

\item \textbf{MS-HGAT} \cite{DBLP:conf/aaai/0004RZLY22} 
builds a sequence of regularly-sampled hypergraphs that contain multiple cascades and users, and then learns on hypergraphs for computing the representations of cascades.

\item \textbf{CasCN} \cite{DBLP:conf/icde/Chen0ZTZZ19} treats each cascade as a graph sequence and uses the GNN and LSTM to learn cascade representations.

\item \textbf{TempCas} \cite{tempcas} additionally designs a sequence modeling method to capture macroscopic temporal patterns apart from learning on the cascade graph. 

\item \textbf{CasFlow} \cite{casflow} is the state-of-the-art method for cascade prediction, which first learns users' representations from the social network and the cascade graph and then employs the GRU and Variational AutoEncoder (VAE) to get representations of cascades. 
\end{itemize}

\begin{table*}[!htbp]
\small
\centering
\label{tab:performance}
\resizebox{0.94\textwidth}{!}{
\setlength{\tabcolsep}{0.9mm}
\begin{tabular}{ccccccccccccc}
\toprule
\multirow{2}{*}{Model} & \multicolumn{4}{c}{Twitter}                                                                               & \multicolumn{4}{c}{Weibo}                                            & \multicolumn{4}{c}{APS}                                               \\ \cmidrule(lr){2-5}  \cmidrule(lr){6-9} \cmidrule(lr){10-13}
                       & MSLE                       & MALE                       & MAPE                      & PCC                         & MSLE            & MALE            & MAPE           & PCC             & MSLE            & MALE            & MAPE           & PCC             \\ \midrule
XGBoost            & 11.5330                    & 2.9871                     & 0.8571                     & 0.3792                      & 3.6253          & 1.3736          & 0.3571          & 0.6493          & 2.5808          & 1.2559          & 0.3437          & 0.4762          \\
MLP            & 11.9105                    & 2.9712                     & 0.9324                     & 0.3733                      & 3.9370          & 1.4409          & 0.3812          & 0.6098          & 2.6075          & 1.2577          & 0.3516          & 0.4787          \\ \midrule
DeepHawkes              & 7.7795                     & 2.1553                     & 0.5547                     & 0.6500                      & 4.2520          & 1.4658          & 0.3998          & 0.5670          & 2.3356          & 1.2001          & 0.3158          & 0.5524          \\
DFTC                   & 5.9173                     & 1.8426                     & 0.4851                     & 0.7495                      & 2.9370          & 1.2046          & 0.2959          & 0.7296          & 2.0357          & 1.1159          & 0.2943          & 0.6247          \\ \midrule
CasCN                  & 7.1021                     & 2.0567                     & 0.5231                     & 0.6940                      & 3.7714          & 1.4040          & 0.3612          & 0.6707          & 2.1248          & 1.1358          & 0.3035          & 0.6062          \\
MS-HGAT                & 5.9992 & 1.9006 & 0.4741 & 0.7507 & OOM             & OOM             & OOM             & OOM             & OOM             & OOM             & OOM             & OOM             \\
TempCas                & 5.5870                      & 1.7584                     & 0.4574                     & 0.7651                      & 2.7453          & 1.1702          & 0.2786          & 0.7500            & 2.0043          & 1.1022          & 0.2957          & 0.6346          \\
CasFlow                & 5.2549                     & 1.5775                     & 0.4031                     & 0.7847                      & 2.6336          & \textbf{1.1230}          & \textbf{0.2687} & 0.7619          & 2.0064          & 1.1053          & 0.2936          & 0.6320          \\
CTCP                   & \textbf{4.6916}            & \textbf{1.5668}            & \textbf{0.3562}            & \textbf{0.8136}             & \textbf{2.5929} & 1.1414 & 0.2723          & \textbf{0.7667} & \textbf{1.6289} & \textbf{0.9906} & \textbf{0.2611} & \textbf{0.7176} \\ \bottomrule
\end{tabular}
}
\caption{Performance of all methods in three datasets, where the methods can be divided into three categories: feature-based, sequence-based, and graph-based methods from top to bottom in the table. The best results appear in bold and OOM indicates the out-of-memory error.}
\label{tab:performance}
\end{table*}

\subsection{Evaluation Metrics}
We choose four widely used metrics to evaluate the performance of the compared methods, including Mean Squared Logarithmic Error (MSLE), Mean Absolute Logarithmic Error (MALE), Mean Absolute Percentage Error (MAPE) and Pearson Correlation Coefficient (PCC). Among these metrics, MSLE, MAPE and MALE evaluate the prediction error between the predicted value and the ground truth from different aspects and PCC measures the correlation between predicted value and the ground truth. 

\subsection{Experimental Settings}\label{sec:experimental_settings}
For XGBoost and MLP, we follow \citet{DBLP:conf/www/ChengADKL14} and extract five types of features (i.e., edge number, max depth, average depth, breath of cascade graph, and publication time of the cascade) as the hand-designed cascade features. We set the dimension of dynamic states of users and cascades, as well as the cascade embedding to 64. The dimension of position embedding is set to 16. The time slot number $n_t$ is set to 20 and the fusion weight $\lambda$ is 0.1. For training, we adopt the Adam optimizer and use the early stopping strategy with a patience of 15. The learning rate and batch size are set to 0.0001 and 50. Our code can be found at \url{https://github.com/lxd99/CTCP}.

\subsection{Performance Comparison}
\tabref{tab:performance} reports the performance of different methods, and some conclusions can be summarized as follows.

Among the three groups of methods, feature-based models perform the worst among all baselines, which reveals that there are complex evolution patterns of the cascade size that can not be captured by the hand-designed features. Moreover, graph-based models show better performance than sequence-based models, implying the necessity of exploiting the structural and temporal information carried in~the~cascade~graph. 

CTCP achieves significant performance improvement w.r.t. the state-of-the-art baseline (i.e., CasFlow) on Twitter and APS, demonstrating the effectiveness of the proposed method. This improvement may be due to the fact that we learn the dynamic representations of cascades and users collaboratively, which can capture the correlation between cascades and the dynamic user preferences outside of a single cascade. The insignificant improvement of CTCP on Weibo may be due to the short time period of Weibo (1 day compared to 1 month and more than
100 years on Twitter and APS respectively) and the preferences of users may not evolve during such a short period, which makes CTCP have no advantages over CasFlow. Additionally, modeling multiple cascades via the sequence of graph snapshots like MS-HGAT does not achieve considerable performance. Because the diffusion behaviors within a snapshot are thought to happen at the same time which will lose fine-grained temporal information. Moreover, MS-HGAT needs to load the snapshot into memory at one time, which makes it can only run on the smallest dataset (i.e., Twitter).

\subsection{Sensitivity to Publication Time}
\begin{figure}[!htbp]
	\begin{subfigure}{0.48\linewidth}
% 		\centering
		\includegraphics[width=\linewidth]{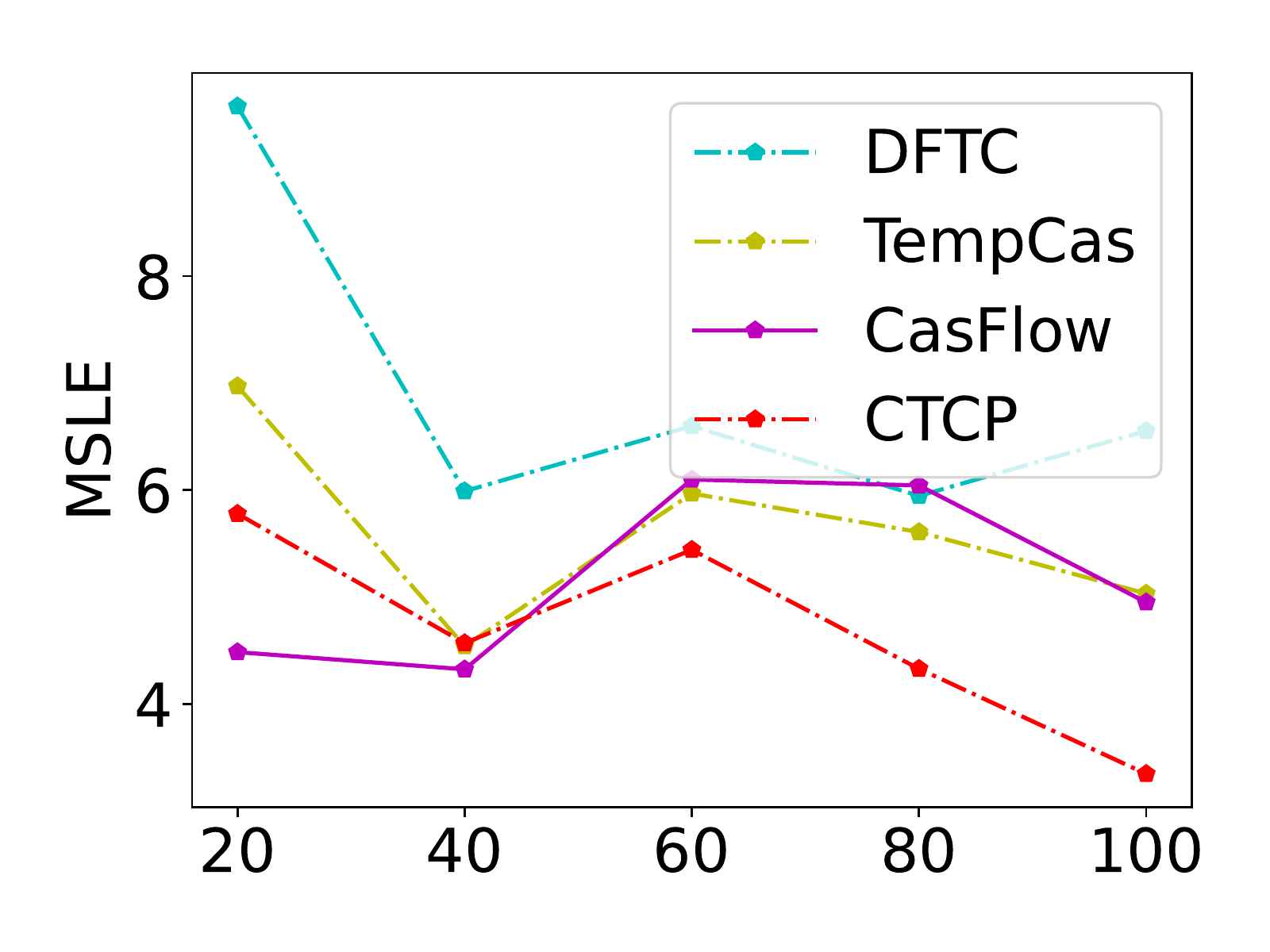}
		\caption{Twitter}
	\end{subfigure}
	\begin{subfigure}{0.495\linewidth}
% 		\centering
		\includegraphics[width=\linewidth]{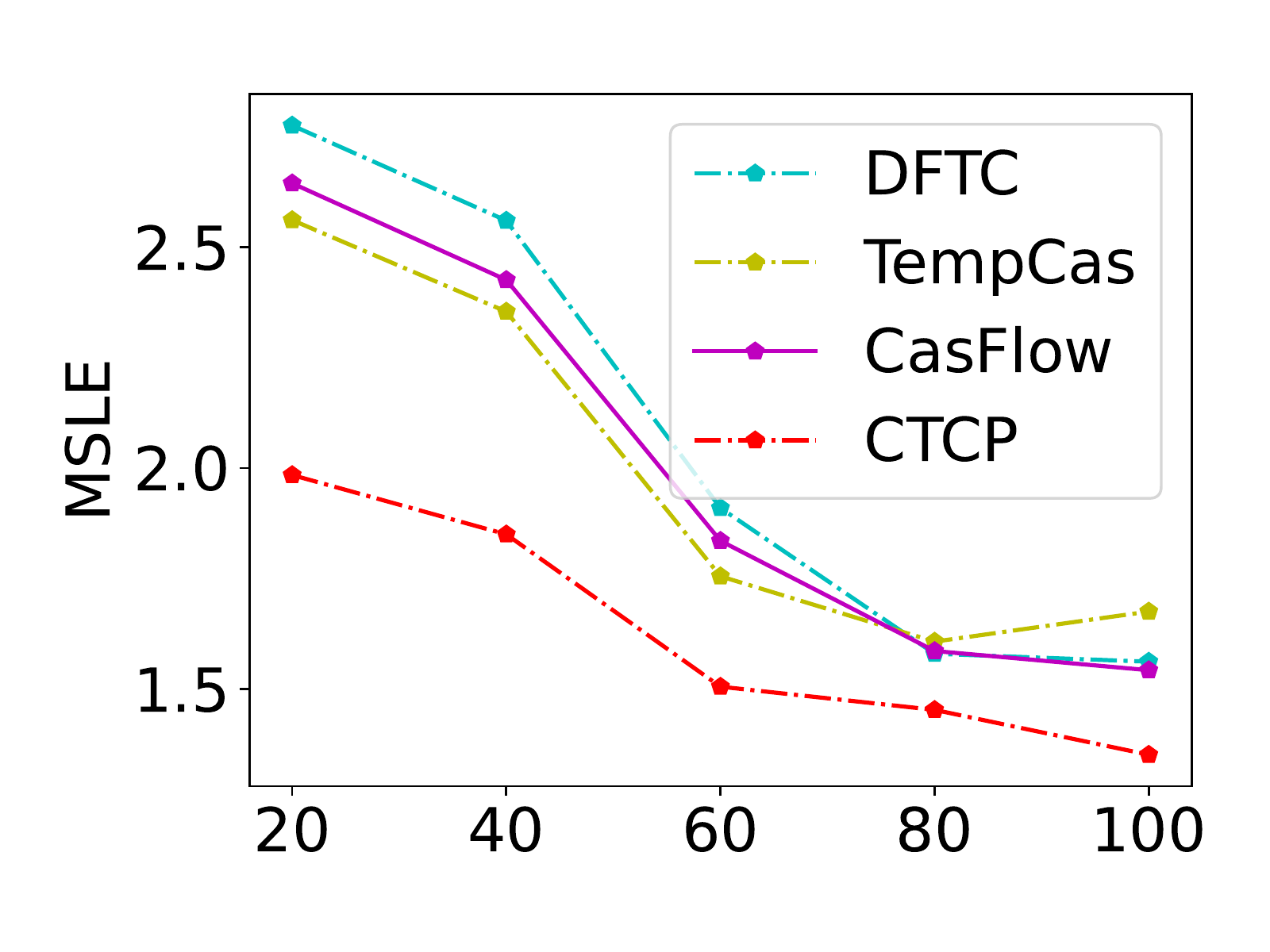}
		\caption{APS}
	\end{subfigure}
	\caption{Performance on cascades with different publication times. The horizontal axis indicates the percentile range of the publication time of cascades and the vertical axis is the average prediction MSLE of cascade in that range.}
	\label{fig:comparison}
\end{figure}

To explore the sensitivity of different models to the publication time of cascades, we plot models' performance on cascades with different publication times on Twitter and APS. Specifically, we divide the cascade into five groups according to their publication time: cascade whose publication time is at the 0th to 20th, 20th to 40th, 40th to 60th, 60th to 80th and 80th to 100th percentile, and plot the best five models' performance.  From \figref{fig:comparison}, we can observe that CTCP can achieve considerable performance on different cascades consistently. Besides, as time goes on, the performance of CTCP consistently improves on these two datasets. This is because the evolution learning module of CTCP keeps updating the dynamic states of users and as time goes on more and more user behaviors are observed, which provides richer information to model the preference of users. Other models only learn from the own diffusion process of cascades and can not learn this dependency.

\subsection{Ablation Study}
\begin{figure}[!htbp]
	\begin{subfigure}{0.485\linewidth}
 		\centering
		\includegraphics[width=\linewidth]{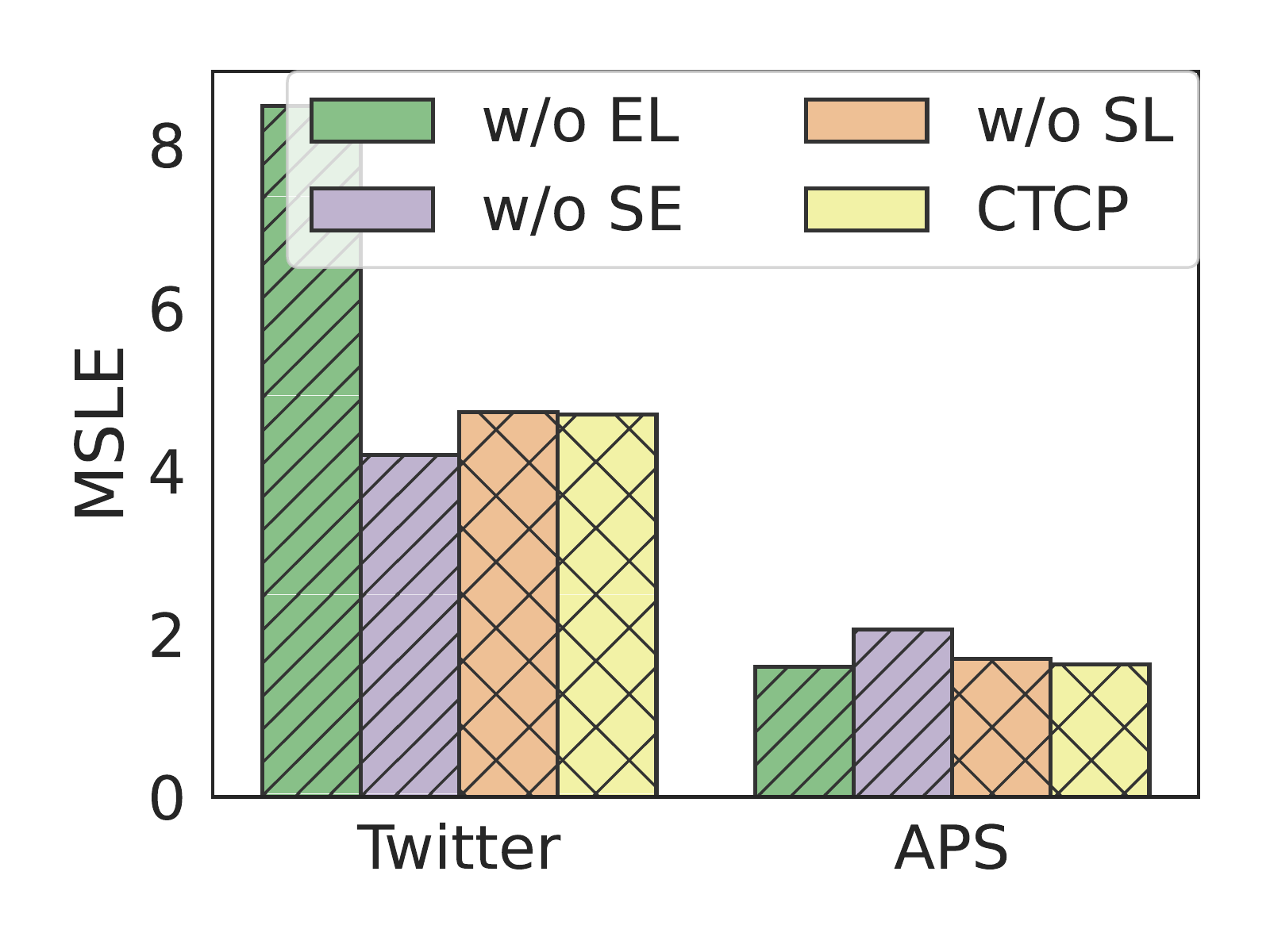}
		\caption{Performance on MSLE}
	\end{subfigure}
	\begin{subfigure}{0.49\linewidth}
		\centering
		\includegraphics[width=\linewidth]{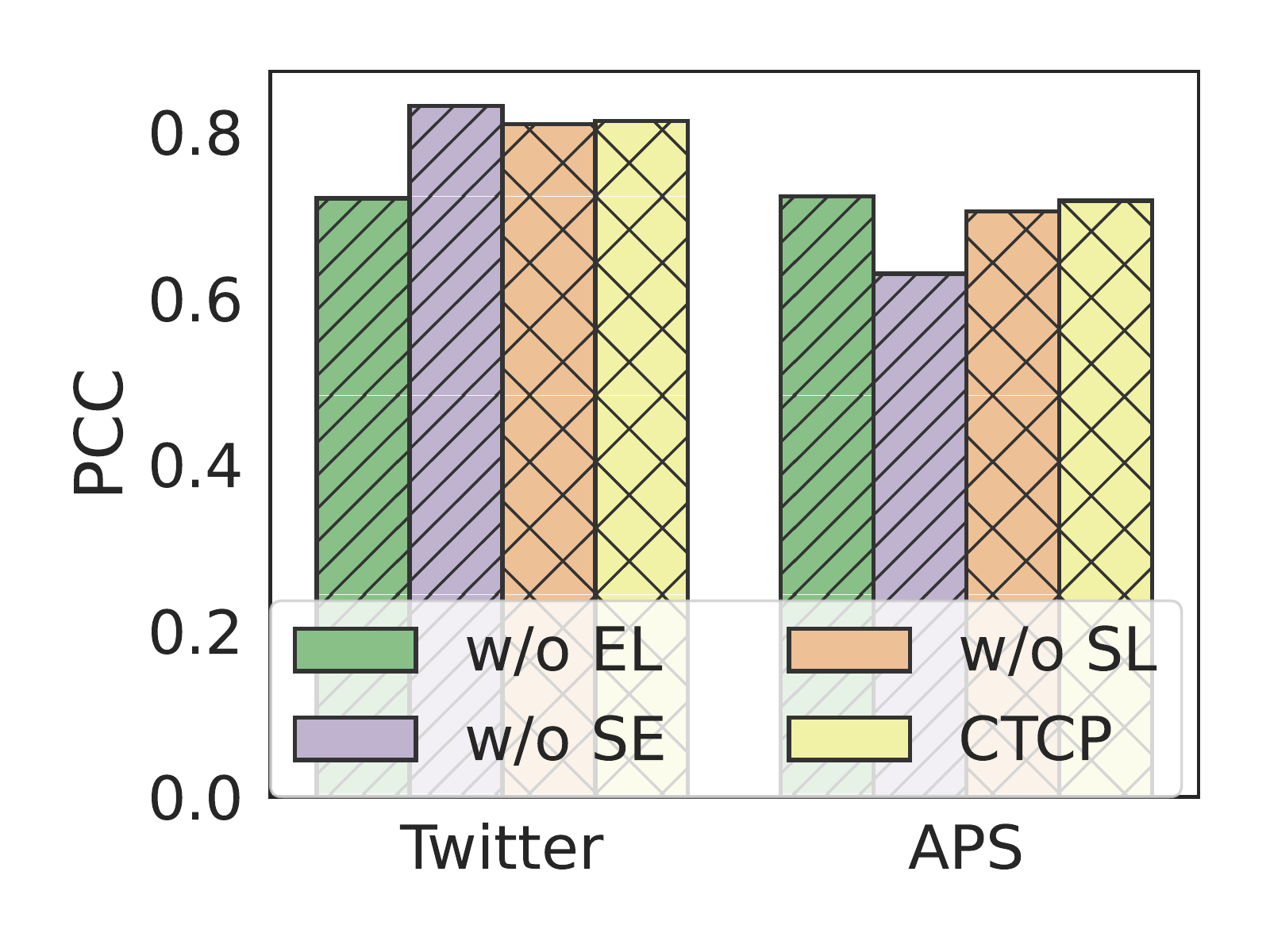}
		\caption{Performance on PCC}
	\end{subfigure}
	\caption{Ablation study on Twitter and APS.}
	\label{fig:ablation}
\end{figure}

We compare CTCP with the following variations on Twitter and APS to investigate the contribution of submodules to the prediction performance.
\begin{itemize}
    \item \textbf{w/o EL} removes the evolution learning module.
    \item \textbf{w/o SE} removes the static representation of users.
    \item \textbf{w/o SL}: removes the structural learning module in the cascade embedding learning process.
\end{itemize}
From \figref{fig:ablation}, we can observe that: Firstly the performance of w/o EL and w/o SE varies on APS and Twitter, for example, w/o SE achieves the best performance on Twitter and the worst performance on APS. This indicates that the growth of the cascade size is controlled by multiple factors and it is necessary to consider the dynamic preference and static preference of users simultaneously.  Secondly, the structural learning module utilizes the cascade graph to generate the cascade embedding which helps improve the prediction
performance by capturing the evolution pattern of a~cascade~at~a~macro~level.

\subsection{Cascade Representations Projection}
\begin{figure}[!htbp]
    \centering
    \includegraphics[width=\linewidth]{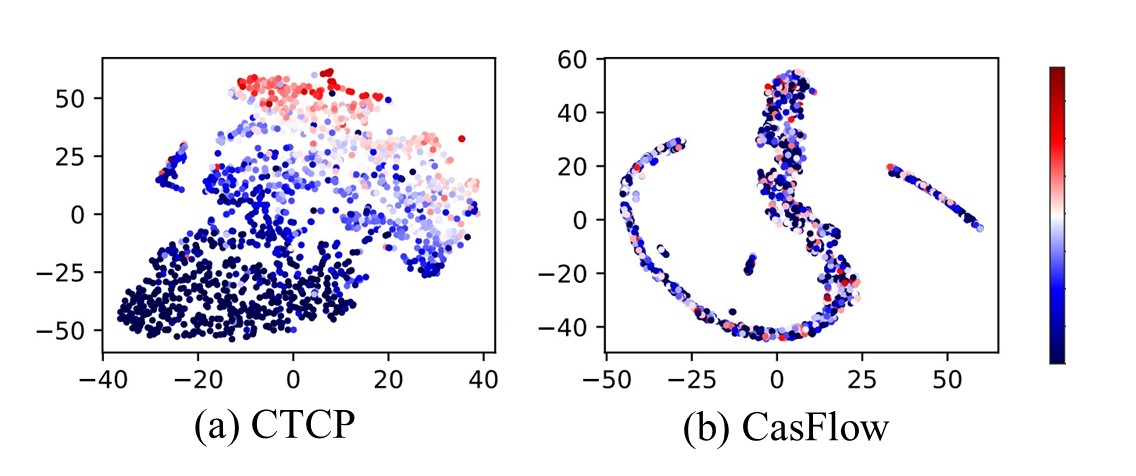}
    \caption{Projection of learned cascade representations on Twitter, where each point represents a cascade representation and the color represents its incremental popularity. (Dark blue means low popularity and dark red means high popularity).}
    \label{fig:projection}
\end{figure}

To confirm the effectiveness of the learned cascade representations, we project the cascade representations of CTCP and CasFlow on Twitter into a two-dimensional space, using t-NSE \cite{JMLR:v9:vandermaaten08a}. Results are represented in \figref{fig:projection}. Remarkably, we find that the learned representations of CTCP can capture the evolution pattern of cascade popularity, suggested by the fact that from right-top to left-bottom the node color of CTCP changes from red to dark blue continuously in \figref{fig:projection} (a). While for CasFlow, nodes with different colors are mixed. This may be because CTCP models the correlation of cascades while CasFlow does not, which can help the model capture the collaborative signals between cascades and learn a better cascade representation.

\section{Conclusion}
In this paper, we studied the problem of cascade popularity prediction and pointed out two factors that are not considered well in the existing methods, i.e., the correlation between cascades and the dynamic preferences of users. Different from previous methods that independently learn from each cascade, our method first combines all cascades into a diffusion graph to explore the correlations between cascades. To model the dynamic preferences of users, an evolution learning module was proposed to learn on the diffusion graph chronologically, which maintains dynamic states for users and cascades, and the states are updated continuously once a diffusion behavior happens.  Moreover, a cascade representation learning module was proposed to explore the structural and temporal information within a cascade by aggregating representations of users into a cascade embedding. Extensive experimental results on three real-world datasets demonstrated the effectiveness of the proposed method.
\section*{Acknowledgements}
The authors would like to thank the anonymous reviewers for their constructive comments on this research. This work was supported by the National Key R$\&$D Program of China (2021YFB2104802) and the National Natural Science Foundation of China (62272023).
\bibliographystyle{named}
\bibliography{ijcai23}
% \subsubsection{Appendices}

\end{document}